\documentclass[aps,prd,reprint,amsmath,amssymb,a4paper]{revtex4-1}

\usepackage[hcentering,vcentering]{geometry}
\usepackage[dvipsnames]{xcolor}
\usepackage[colorlinks=true,allcolors=Blue,breaklinks=true]{hyperref}

\usepackage{multirow}

\usepackage{newtxtext,newtxmath}

\usepackage{graphicx}

\usepackage{enumitem}

\newcommand{\Mp}{\ensuremath{M_\text{Pl}}}

\begin{document}

\title{One-electron atoms in screened modified gravity}

\author{Leong Khim \surname{Wong}}
\email[]{L.K.Wong@damtp.cam.ac.uk}
\author{Anne-Christine \surname{Davis}}
\email[]{A.C.Davis@damtp.cam.ac.uk}
\affiliation{Department of Applied Mathematics and Theoretical Physics, Centre for Mathematical Sciences, University of Cambridge, Wilberforce Road, CB3 0WA, United Kingdom}

\date{May 9, 2017}

\begin{abstract}
In a large class of scalar-tensor theories that are potential candidates for dark energy, a nonminimal coupling between the scalar and the photon is possible. The presence of such an interaction grants us the exciting prospect of directly observing dark sector phenomenology in the electromagnetic spectrum. This paper investigates the behavior of one-electron atoms in this class of modified gravity models, exploring their viability as probes of deviations from general relativity in both laboratory and astrophysical settings. Building heavily on earlier studies, our main contribution is threefold: A thorough analysis finds additional fine-structure corrections previously unaccounted for, which now predict a contribution to the Lamb shift that is larger by nearly 4 orders of magnitude. In addition, they also predict a scalar-mediated photon-photon interaction, which now constrains the scalar's coupling to the photon independently of the matter coupling. This was not previously possible with atomic precision tests. Our updated constraints are $\log_{10}\beta_m \lesssim 13.4$ and $\log_{10}\beta_\gamma \lesssim 19.0$ for the matter and photon coupling, respectively, although these remain uncompetitive with bounds from other experiments. Second, we include the effects of the nuclear magnetic moment, allowing for the study of hyperfine structure and the 21\,cm line, which hitherto have been unexplored in this context. Finally, we also examine how a background scalar leads to equivalence principle violations.
\end{abstract}

\maketitle

\section{Introduction}

Scalar fields have become ubiquitous in modern theoretical physics, playing a crucial role as potential solutions for many of the most important open problems. They have been posited as the driving force behind inflation, as a solution to the strong-\emph{CP} problem, and as candidates for dark matter and dark energy. Scalar fields are also interesting from the perspective of string theory, where scalar-tensor gravity actions naturally arise after dimensional reduction.

Despite many instances in which the existence of a fundamental scalar proves desirable, none---apart from the Higgs boson---has been observed to date. For those with applications to late-time cosmology, the resolution comes by way of introducing a screening mechanism. This enables the scalar to evade the stringent constraints of fifth force searches~\cite{GRtest} while still potentially giving rise to astrophysical signatures. Of particular interest in this work are classes of models where the scalar has a canonical kinetic term in the Einstein frame, with screening prompted by a nonminimal coupling to matter. These include screening of the chameleon~\cite{Cham1,*Cham2} and Damour-Polyakov types~\cite{DamourPolyakov,EDDilaton,Symmetron,*SymmetronCosm}. It is interesting to also allow for a non-minimal coupling between the scalar and the photon since one is not forbidden by symmetry. This is reminiscent of Bekentein-Sandvik-Barrow-Magueijo varying alpha models~\cite{Bekenstein,*SBM,*GeneralizedBSBM}, the Olive-Pospelov model~\cite{OlivePospelov}, Kaluza-Klein theories, and the dilaton in string theory. Its phenomenology in the present context has been widely explored, both in astronomical~\cite{ChamCMB,SNeBrightening,SolarChameleon1,*SolarChameleon2,*SolarChameleon3,CAST,SymmetronSpatialDipole,ChamPolarization} and laboratory settings~\cite{ChamBirefringence,ChamPVLAS,Alpenglow,GammeV,GammeVCHASE}.

In this paper, we ask if atomic spectra can be used as a tracer for modifications to gravity. The prototypical candidate for such a task is hydrogen, for two reasons: Hydrogen is readily abundant in the Universe, and its spectral lines find important uses in inferring galactic rotation curves~\cite{Rubin,*Bosma,SPARC} and as probes of large-scale structure and cosmic history~\cite{HIntensityMapping1,*HIntensityMapping2}. It is therefore important that any systematic effect coming from modified gravity is understood. Second, its relative simplicity enables us to obtain analytic solutions using standard methods of perturbation theory.

Hydrogen has already enjoyed an illustrious history as a theoretical playground for probing gravitational effects. The effect of tidal forces from a strongly curved background was investigated in~\cite{Parker,ParkerPimentel}, with further calculations specific to the Schwarzschild metric done in~\cite{Gill}. Hydrogen atoms have also been studied near topological defects~\cite{HTopDefects}, in both de~Sitter and anti-de~Sitter space~\cite{hdS,*HAdS}, in Rindler space~\cite{HRindler}, and in $f(R)$ theories with low curvature corrections~\cite{hfR}. In fact, one-electron atoms in chameleonlike scalar-tensor theories have already been explored in~\cite{BraxBurrage}. However, the implications of the scalar-photon coupling were not fully exhausted, and we do so by extending their findings in this paper.

The outline of this paper is as follows: The class of scalar-tensor theories under consideration is briefly reviewed in Sec.~\ref{sec:ScalarTensor}. We enumerate all the possible gravitational effects that lead to first-order perturbations in the Hamiltonian in Sec.~\ref{sec:ClassicalFields}, and the Hamiltonian itself is derived in Sec.~\ref{sec:H}. Its implications are discussed in Sec.~\ref{sec:Discussion}, and we conclude in Sec.~\ref{sec:Conclusion}.

\section{Scalar-tensor gravity}
\label{sec:ScalarTensor}

We consider a scalar-tensor theory of gravity with the action
\begin{equation}
S = \int \text{d}^4x\; (\mathcal L_\text{grav} + \mathcal L_\text{em} + \mathcal L_\text{m}).
\end{equation}
The gravitational sector in the Einstein frame has the Lagrangian
\begin{equation}
\frac{\mathcal L_\text{grav}}{\sqrt{-\tilde g}} = \frac{\Mp^2}{2}(\tilde R - \tilde g^{\mu\nu} \partial_\mu\phi\,\partial_\nu\phi) - V(\phi),
\label{eq:LagrangianGravEinstein}
\end{equation}
where $\Mp = (8\pi G_\text{N})^{-1/2}$ is the reduced Planck mass in units $\hbar = c = 1$. Our metric signature is $(-+++)$ throughout. The scalar couples nonminimally to the electromagnetic sector~\cite{AnomCouplings,HiggsProduction}, such that
\begin{equation}
\frac{\mathcal L_\text{em}}{\sqrt{-\tilde g}} = -\frac{1}{4} \varepsilon(\phi) F_{\mu\nu} F^{\mu\nu}.
\label{eq:LagrangianEM}
\end{equation}
The matter sector, described by $\mathcal L_\text{m}$, couples minimally to the Jordan frame metric $g_{\mu\nu} = \Omega^2(\phi) \tilde g_{\mu\nu}$. A scalar-tensor theory of this class is fully specified by a choice of the three functions $V(\phi)$, $\varepsilon(\phi)$, and $\Omega(\phi)$. The results of this paper hold for any theory with the above action, although we will henceforth refer to all such scalars generically as chameleons.

The primary aim of this work is to extend the findings of~\cite{BraxBurrage} by including all of the effects due to a chameleon-photon coupling at first order in perturbations. Only a subset of these were previously accounted for. Furthermore, as the calculations in~\cite{BraxBurrage} were made in the Einstein frame, we have elected to work in the Jordan frame. This gives us, as an added bonus, a concrete example with which to make interesting comparisons between the two approaches (see Sec.~\ref{sec:Frames}). In the Jordan frame, we have
\begin{equation}
\frac{\mathcal L_\text{grav}}{\sqrt{-g}} = \frac{1}{2}\Mp^2\Omega^{-2} \left[R - (1-6\beta_m^2)(\partial\phi)^2 - U \right].
\end{equation}
We have introduced the Jordan-frame potential $U=2\Mp^{-2}\Omega^{-2}V$ to simplify several equations, but we will continue to use both $U$ and $V$. The function $\beta_m(\phi)$ describes the coupling of the chameleon to matter and is defined as $\beta_m = (\log\Omega)_{,\phi}$, where a comma denotes a partial derivative. As the electromagnetic sector is (classically) conformally invariant, its Lagrangian in the Jordan frame is identical to that in Eq.~\eqref{eq:LagrangianEM}, except with $\tilde g_{\mu\nu}$ replaced with $g_{\mu\nu}$. Similarly, let $\beta_\gamma = (\log\varepsilon)_{,\phi}$ describe the chameleon-photon coupling. The equations of motion in this frame are
\begin{subequations}
\label{eq:eom}
\begin{align}
G_{\mu\nu} &= \frac{\Omega^2}{\Mp^2}T_{\mu\nu} + T^{(\phi)}_{\mu\nu},
\label{eq:eomGrav}
\\
\Box\phi - 2\beta_m(\partial\phi)^2 &= \frac{\Omega^2}{\Mp^2}
\left( \frac{V_{,\phi}}{\Omega^4} - \beta_m T + \frac{\varepsilon \beta_\gamma}{4}  F^2 \right),
\label{eq:eomCham}
\\
\nabla_\mu(\varepsilon F^{\mu\nu}) &= - J^\mu,
\label{eq:eomEm}
\end{align}
\end{subequations}
where $F^2 \equiv F_{\mu\nu}F^{\mu\nu}$, $J^\mu$ is the total conserved current charged under the U(1) gauge group, and $T_{\mu\nu}$ is the stress tensor of all matter fields, with trace $T$. This includes the contribution of the electromagnetic stress tensor, whose form is slightly modified due to the nonminimal coupling:
\begin{equation}
T^\text{(em)}_{\mu\nu} = \varepsilon(\phi) \left(F_{\mu\rho} F_\nu{}^\rho - \frac{1}{4} g_{\mu\nu} F^2 \right).
\end{equation}
The stress tensor for the chameleon is
\begin{align}
T_{\mu\nu}^{(\phi)} = &\;
2\beta_m(g_{\mu\nu}\Box\phi-\nabla_\mu\nabla_\nu\phi)
\nonumber\\
&- \frac{1}{2}\left[(1 - 4\beta_{m,\phi} + 2\beta_m^2)(\partial\phi)^2 + U \right]g_{\mu\nu}
\nonumber\\
&+ (1 - 2\beta_{m,\phi} - 2\beta_m^2)\partial_\mu\phi\,\partial_\nu\phi.
\end{align}

\section{Classical field solutions}
\label{sec:ClassicalFields}

Our model of the atom is semiclassical: The nucleus sources gravitational, electromagnetic, and chameleon fields, on which we wish to quantize the electron. This section discusses their solutions in turn.

\subsection{Background curvature}
\label{sec:Background}

As we are interested in the, admittedly small, gravitational effects on atoms, we should wonder if the background spacetime causes an appreciable perturbation. For a point-particle nucleus traversing the worldline $\Gamma$ in a background spacetime $\bar g_{\mu\nu}$, there always exists adapted coordinates $(t,x^i)$ such that the rest frame of the nucleus can be Taylor expanded to give
\begin{equation}
\bar g_{\mu\nu} = \eta_{\mu\nu} + \bar g_{\mu\nu,i} x^i + \frac{1}{2} \bar g_{\mu\nu,ij} x^i x^j + \mathcal O(r^3),
\end{equation}
where all derivatives are evaluated at the spatial origin $x^i = 0$ but still generally remain functions of time $t$. While the choice of coordinates $(t,x^i)$ is by no means unique, the Fermi coordinates are especially useful~\cite{ManasseMisner,Poisson2011}. These preserve the notion of proper time $t$ along the entire extent of the worldline $\Gamma$ and the notion of proper distances $x^i$ in its convex normal neighborhood. At the order prescribed, two types of terms appear: terms linear in the background Riemann tensor $\bar R_{\mu\nu\rho\sigma}$, which describe tidal forces, and terms up to quadratic order in the acceleration $a^\mu$ of $\Gamma$. The interested reader can find explicit expressions for the metric components in~\cite{Poisson2011}.

The effect of the Riemann tensor terms are studied in~\cite{Parker}, with the result that the leading contribution to the Hamiltonian is
\begin{equation}
H \supset \frac{1}{2} m_e \bar R_{0i0j} x^i x^j \beta,
\end{equation}
where $\beta$ is a Dirac matrix as defined in Sec.~\ref{sec:H}. Making an order of magnitude estimate and substituting $\bar R_{0i0j} \sim \mathcal{D}^{-2}$ with some curvature length scale $\mathcal D$, this perturbation is comparable to fine-structure splitting only when $\mathcal D \lesssim Z^{-3}\,10^{-4}\,\text{cm}$, where $Z$ is the atomic number. Galactic (stellar mass) black holes, assumed to be some of the most strongly curved regions of spacetime in our Universe, have $\mathcal D \sim 10^5\,\text{cm}$ near its event horizon. On these length scales, the perturbation would give rise to an energy shift $\sim 10^{-24}\,\text{eV}$; on Earth it would be $10^{-38}\,\text{eV}$. This is unobservable in both cases.

Similarly, the effect of an accelerating worldline is studied in~\cite{HRindler}, with the leading contribution being
\begin{equation}
H \supset m_e a_i x^i \beta.
\end{equation}
If we again wish for this to be comparable to fine-structure splitting, this requires $a \gtrsim 10^{18}\,g_\oplus$, where $g_\oplus$ is the Earth's surface gravity. On Earth, this perturbation would give rise to energy shifts $\sim 10^{-21}\,\text{eV}$. While consistently a larger effect than tidal forces, this is also unobservable. Consequently, contributions from a curved background can be safely ignored in what follows.

\subsection{Chameleon profile}
\label{sec:ChamProfile}

In the rest frame of the nucleus, a background chameleon can be Taylor expanded to give $\phi^{(0)} = \bar\phi + \bar\phi_i x^i + \mathcal O(r^2)$. The constant $\bar\phi$ is the chameleon's vacuum expectation value (vev) in the neighborhood of the atom, and $\bar\phi_i \equiv \partial_i \bar\phi|_{x=0}$ is a background gradient. Overlaying this background is a local profile $\delta\phi = \phi - \phi^{(0)}$ due to the presence of the nucleus. Linearizing Eq.~\eqref{eq:eomCham} about $\phi^{(0)}$, and considering only static solutions, we obtain
\begin{equation}
\nabla^2 \delta\phi = \frac{\bar\Omega^2}{\Mp^2}\left[ \bar\beta_m m_N \delta^{(3)}({\bf r}) - \frac{1}{2} \bar\varepsilon \bar\beta_\gamma {\bf E}^2_{(0)} \right],
\label{eq:eomChamLinear}
\end{equation}
where $\nabla^2 \equiv \delta^{ij}\partial_i\partial_j$ is the flat-space Laplacian, and overbars denote a quantity evaluated at $\bar\phi$, e.g., $\bar\Omega \equiv \Omega(\bar\phi)$. This equation retains only the two dominant sources for $\delta\phi$, the mass $m_N$ and the Coulomb field ${\bf E}_{(0)}$ of the nucleus. Several subleading terms have been safely discarded:
\begin{enumerate}[itemsep=0pt,topsep=0pt,parsep=0pt,label={(\arabic*)}]
\item On the lhs, we ignore the kinetic term $\bar\phi^i \partial_i\delta\phi$ which describes a local-background chameleon interaction. If the background changes on a length scale $\mathcal D$, and the Bohr radius is $a_0$, then this term is suppressed by at least a factor $a_0/\mathcal D \ll 1$ relative to $\nabla^2\delta\phi$.

\item On the rhs, we have ignored the magnetic moment of the nucleus. The resulting magnetic field ${\bf B}_{(0)}$ is always several orders of magnitude weaker than the Coulomb field. Note that excluding the magnetic contribution here does not preclude us from studying the leading chameleon effect on hyperfine structure in Sec.~\ref{sec:hfs}.

\item We have also ignored any large-scale, background electromagnetic fields. Relaxing this assumption would give rise to two types of terms. The first are purely background terms $\sim {\bf E}^2_\text{(bg)}$, ${\bf B}^2_\text{(bg)}$ which source the background chameleon $\phi^{(0)}$~\footnote{Note that electromagnetic \emph{waves} are special in that their contribution $F^2 = 2({\bf B}^2-{\bf E}^2)$ vanishes, so they have no effect altogether on the background chameleon.}, having no bearing on the form of $\delta\phi$. The second type are cross terms proportional to ${\bf E}_{(0)} \cdot {\bf E}_\text{(bg)}$ and ${\bf B}_{(0)} \cdot {\bf B}_\text{(bg)}$. The conclusions of this paper hold provided $|{\bf E}_\text{(bg)}|, |{\bf B}_\text{(bg)}| \ll |{\bf E}_{(0)}|$ in the vicinity of the nucleus, which is typical for environments that host atoms, rather than ions.

\item Finally, we have also ignored the term $m_\phi^2\delta\phi$ that accounts for the chameleon's effective mass in this environment,
\begin{equation}
m^2_\phi = \frac{1}{\Mp^2} \frac{\text{d}}{\text{d}\phi}
\left.\left( \frac{V_{,\phi}}{\Omega^2} - \beta_m \Omega^2 T^{(0)} \right)\right|_{\bar\phi},
\end{equation}
where $T^{(0)}$ is the background matter density. This expression omits background electromagnetic fields, whose effect is typically small compared with the matter density. We can neglect this term if the local chameleon's Compton wavelength $m_\phi^{-1}$ is much larger than the Bohr radius, that is, $m_\phi \ll \zeta m_e \approx Z \, (3.7\,\text{keV})$. This is easily satisfied by most, if not all, screening mechanisms of physical interest. Of course, our results equally apply to unscreened scalar fields with mass below this upper limit.
\end{enumerate}

Taking the electric field to be $|{\bf E}_{(0)}| = Ze/4\pi\bar\varepsilon r^2$ (see Sec.~\ref{sec:Maxwell}), the solution is
\begin{equation}
\delta\phi = - \frac{\bar\Omega^2}{8\pi\Mp^2}
\left( \frac{2\bar\beta_m m_N}{r} + \frac{\bar\beta_\gamma Z^2 \alpha}{2\bar\varepsilon r^2} \right).
\label{eq:ChamSolution}
\end{equation}
In what follows, one finds that $\bar\Omega^2$ always accompanies a factor of $(8\pi\Mp^2)^{-1}$, and $\bar\varepsilon^{-1}$ always accompanies the fine-structure constant $\alpha$. The reason for this is clear: The effect of $\bar\Omega$ is to induce an effective gravitational constant, while $\bar\varepsilon$ induces an effective fine-structure constant. For brevity, we will henceforth write
\begin{equation}
G = \frac{\bar\Omega^2}{8\pi\Mp^2} = \bar\Omega^2 G_\text{N},\quad
\zeta = \frac{Z\alpha}{\bar\varepsilon}.
\label{eq:VaryingConstants}
\end{equation}
We will also drop the overbars on $\beta_m$ and $\beta_\gamma$.

\subsection{Metric perturbations}
This local chameleon field and the mass of the nucleus go on to source a metric perturbation $h_{\mu\nu} = g_{\mu\nu} - \eta_{\mu\nu}$. Linearizing Eq.~\eqref{eq:eomGrav}, we find
\begin{equation}
\delta G_{\mu\nu} = 8\pi G \delta T_{\mu\nu}
+ 2\beta_m(\eta_{\mu\nu}\nabla^2\delta\phi - \partial_\mu \partial_\nu \delta\phi).
\label{eq:eomGravLinear}
\end{equation}
The stress tensor of the nucleus is $\delta T_{\mu\nu} = m_N \delta^{(3)}({\bf r}) \delta_\mu^0\delta_\nu^0$, and the linearized Einstein tensor is
\begin{equation}
2\delta G_{\mu\nu} =
-\Box \hat h_{\mu\nu} - \eta_{\mu\nu} \partial^\alpha\partial^\beta \hat h_{\alpha\beta}
+ 2\partial^\alpha \partial_{(\mu} \hat h_{\nu)\alpha},
\end{equation}
where $\hat h_{\mu\nu}$ is the trace-reversed metric perturbation. For similar reasons, we have ignored the term that is of the order $\sim m_\phi^2 \delta\phi \, \eta_{\mu\nu}$ on the rhs of Eq.~\eqref{eq:eomGravLinear}.

It is easy to verify that this has the solution
\begin{equation}
h_{\mu\nu} = -2 \Phi_\text{N} \delta_{\mu\nu} + 2 \Phi_\text{S} \eta_{\mu\nu},
\end{equation}
where the first term is the familiar weak-field metric with Newtonian potential $\Phi_\text{N} = - G m_N/r$, and the second is a fifth force potential $\Phi_\text{S} = \beta_m \delta\phi$. Defining $\Phi_\pm = \Phi_\text{N} \pm \Phi_\text{S}$, the metric can be written as
\begin{equation}
\text{d}s^2 = -(1+2\Phi_+)\, \text{d}t^2 + (1-2\Phi_-)\, \text{d}{\bf x}^2.
\label{eq:WeakFieldMetric}
\end{equation}
The nonvanishing components of the Christoffel symbols are
\begin{align}
\Gamma^0{}_{0i} &= \partial_i \Phi_+,\quad
\Gamma^i{}_{00} = \partial^i \Phi_+,
\nonumber\\
\Gamma^k{}_{ij} &= \partial^k \Phi_- \delta_{ij} - 2 \delta^k_{(i}\partial_{j)}\Phi_-.
\end{align}

It will later prove useful to express the metric in terms of the vierbeins $\{ e_\mu^{\hat a} \}$, $g_{\mu\nu} = \eta_{\hat a \hat b} e^{\hat a}_\mu e^{\hat b}_\nu$, where hatted Roman indices $\hat a, \hat b$ denote local Lorentz indices. When the distinction is unnecessary, the hats will be dropped. Their dual vector fields are defined as $e^\mu_{\hat a} = \eta_{\hat a \hat b} g^{\mu\nu} e^{\hat b}_\nu$. For the weak-field metric in Eq.~\eqref{eq:WeakFieldMetric}, these are
\begin{align}
e_\mu^{\hat 0} = (1+ \Phi_+) \delta_\mu^0,
\quad &
e_\mu^{\hat\imath} = (1-\Phi_-) \delta_\mu^i,
\nonumber\\
e^\mu_{\hat 0} = (1  - \Phi_+) \delta^\mu_0,
\quad &
e^\mu_{\hat\imath} = (1+\Phi_-)\delta^\mu_i.
\end{align}
Also useful are expressions for the Ricci tensor, whose non-vanishing components are
\begin{equation}
R_{00} = \nabla^2\Phi_+, \quad
R_{ij} = \delta_{ij} \nabla^2\Phi_- - 2\partial_i\partial_j\Phi_\text{S}.
\end{equation}

\subsection{Maxwell equations}
\label{sec:Maxwell}

In the Lorenz gauge $\nabla^\mu A_\mu = 0$, Eq.~\eqref{eq:eomEm} can be rewritten as
\begin{equation}
\Box A_\mu - R_\mu{}^\nu A_\nu - \beta_\gamma F_{\mu\nu} \nabla^\nu \phi = - \frac{J_\mu}{\varepsilon},
\label{eq:eomMaxwell}
\end{equation}
which must be solved on the weak-field metric to consistently keep all terms linear in $\phi$. Ignoring any large-scale, background electromagnetic fields as before, we expand $A_\mu = A_\mu^{(0)} + \delta A_\mu$, where $A_\mu^{(0)}$ describes the bare electric and magnetic fields of the nucleus,
\begin{equation}
eA_0^{(0)} = - \frac{\zeta}{r}, \quad
eA_i^{(0)} = \gamma \frac{\epsilon_{ijk} I^j x^k}{r^3},
\label{eq:BareGaugeField}
\end{equation}
sourced by its charge and magnetic moment, respectively. The constant $\zeta = Z\alpha/\bar\varepsilon$ is as defined in Sec.~\ref{sec:ChamProfile}, $I^j$~is the spin operator of the nucleus, and $\gamma = g\alpha/2 \bar\varepsilon m_p$ is proportional to the nuclear gyromagnetic ratio~\footnote{Precisely, the nuclear gyromagnetic ratio multiplied by $e/4\pi\bar\varepsilon$.}.

Linearizing Eq.~\eqref{eq:eomMaxwell}, we obtain
\begin{subequations}
\begin{align}
\nabla^2 \delta A_0 =&\,
\beta_\gamma(\bar\phi^i + \partial^i\delta\phi)F_{0i}^{(0)} + 2\partial_i \Phi_\text{N}\partial^i A_0^{(0)},
\label{eq:eomMaxwellA0}
\\[0.5em]
\nabla^2 \delta A_m =&\,
\beta_\gamma(\bar\phi^i + \partial^i\delta\phi) F_{mi}^{(0)} - 2 \partial_i \Phi_\text{N} \partial^i A_m^{(0)}
\nonumber\\
&+ 2 \partial^i\Phi_- \partial_m A_i^{(0)} - 2 A_i^{(0)} \partial^i\partial_m \Phi_\text{S}.
\end{align}
\end{subequations}
These equations admit the natural interpretation that gravitational effects generate secondary charges and currents, which then source corrections to the bare electromagnetic fields. The appropriate boundary condition for $\delta A_\mu$ is, therefore, that it vanishes in the absence of these effects. Said differently, we demand that the complementary functions be zero. For the zeroth-order gauge fields as given in Eq.~\eqref{eq:BareGaugeField}, the solutions are
\begin{subequations}
\label{eq:MaxwellSolutions}
\begin{align}
e \delta A_0 =&\,
\frac{\zeta\beta_\gamma\bar\phi_i x^i}{2r} - \frac{G\zeta(\beta_m\beta_\gamma - 1) m_N}{r^2}
\nonumber\\
&- \frac{G\zeta^2 \beta_\gamma^2 Z}{6 r^3},
\\
e \delta A_m =&\,
\gamma\beta_\gamma \epsilon_{ik[m}\delta_{l]j}\bar\phi^l \frac{x^i x^j I^k}{r^3}
\nonumber\\
&- \frac{\gamma}{2}G(\beta_m \beta_\gamma + 1)m_N \epsilon_{ijm} \frac{x^i I^j}{r^4}
\nonumber\\
&-\frac{\gamma}{10}G Z\zeta \beta_\gamma^2\epsilon_{ijm}\frac{x^i I^j}{r^5}.
\end{align}
\end{subequations}
The Lorenz gauge at first order is
\begin{equation}
\partial^\mu \delta A_\mu = - 2 A_i^{(0)} \partial^i \Phi_\text{S},
\end{equation}
and we have verified that our solutions satisfy this condition.

\section{Hamiltonian}
\label{sec:H}

In the previous section, we obtained classical solutions for the gravitational, electromagnetic, and chameleon field profiles sourced by a nonrelativistic nucleus fixed at the origin. This section derives the Hamiltonian for an electron moving in such a background. In the Jordan frame, the chameleon couples to the electron only indirectly through its effect on the metric and gauge fields, making the Dirac equation the natural starting point,
\begin{equation}
(\underline\gamma^\mu D_\mu + m_e) \psi = 0.
\label{eq:Dirac}
\end{equation}
It is certainly within our prerogative to take the nonrelativistic limit once we have obtained the Hamiltonian, as is sometimes done, but this comes with little advantage. Working with the relativistic wave functions is only marginally more involved, whereas a nonrelativistic expansion in orders of $\alpha$ generate an unnecessarily long list of terms to deal with. For these reasons, our calculations are kept relativistic throughout.

Our conventions follow mostly those of~\cite{Parker}, where the position-dependent gamma matrices satisfy $\{ \underline\gamma^\mu, \underline\gamma^\nu \} = 2 g^{\mu\nu}$. We can of course write $\underline\gamma^\mu(x) = e^\mu_a(x)\,\gamma^a$, where $\gamma^a$ are the flat-space variants. The electron has charge $-e$, and thus has covariant derivative $D_\mu = \partial_\mu - \omega_\mu + ieA_\mu$,
with spin connection
\begin{equation}
\omega_\mu = \frac{1}{2}\gamma_{ab} (\partial_\mu e_\nu^a - \Gamma^\alpha{}_{\nu\mu} e_\alpha^a)g^{\nu\lambda} e^b_\lambda.
\end{equation}
The matrices $\gamma_{ab} = \frac{1}{4}[\gamma_a,\gamma_b]$ are the generators of the Lorentz algebra. For the weak-field metric in Eq.~\eqref{eq:WeakFieldMetric}, this evaluates to
\begin{equation}
\omega_0 = \gamma^{0i}\partial_i \Phi_+,\quad
\omega_i = \gamma_{ij} \partial^j\Phi_-.
\end{equation}

\begin{table*}
\caption{Additive fine-structure corrections from chameleon effects to the energies of three transition lines: the Ly\,$\alpha$ line in electronic hydrogen, and the Lamb shifts in electronic and muonic hydrogen. We have listed the contributions from each of the different terms in Eq.~\eqref{eq:HChmFsExp}. The numerical values given assume $\bar\Omega = \bar\varepsilon = 1$, appropriate for a laboratory on Earth. The three rightmost columns give the standard uncertainties for the energy levels as predicted by standard QED theory, and for the experimental measurement of the transition frequency.}
\label{table:fs}
\begin{ruledtabular}
\begin{tabular}{cccccccc}
& \multicolumn{4}{c}{Additive corrections from chameleon effects (eV)}
& \multicolumn{3}{c}{Standard uncertainties%
\footnote{%
These values were obtained from Table~XVI of~\cite{Codata}. Note that $1\,\text{kHz} = 4.1\times10^{-12}\,\text{eV}$. The third row is not used to constrain $\beta_m$ and $\beta_\gamma$, so no values are given.
}
(kHz)}
\\[0.2em]
\cline{2-5}\cline{6-8}
Transition & $\mathcal B_{-1}$ term & $\mathcal B_{-2}$ term & $\mathcal A_{-2}$ term & $\mathcal A_{-3}$ term 
& $\sigma(\psi)_\text{th}$ & $\sigma(\psi')_\text{th}$ & $\sigma(\psi-\psi')_\text{exp}$  \\[0.2em]
\hline
$2S_{1/2}-1S_{1/2}\;(e^-)\;\,$
& $1.8\times 10^{-38} \,\beta_m^2$
& $3.0\times 10^{-46} \,\beta_m\beta_\gamma$
& $1.1\times 10^{-42} \,\beta_m\beta_\gamma$
& $1.1\times 10^{-49} \,\beta_\gamma^2$
& 0.31
& 2.5
& 0.010
\\
$2S_{1/2}-2P_{1/2} \;(e^-)\;\,$
& $0$
& $2.9\times 10^{-47} \,\beta_m\beta_\gamma$
& $1.1\times 10^{-43} \,\beta_m\beta_\gamma$
& $1.5\times 10^{-50} \,\beta_\gamma^2$
& 0.31
& 0.028
& 9.0
\\
$2P_{1/2}-2S_{1/2} \;(\mu^-)$\footnote{Note that the $2P_{1/2}$ level is raised relative to the $2S_{1/2}$ level in muonic hydrogen, while the converse is true in electronic hydrogen.}
& $0$
& $2.6\times 10^{-40} \,\beta_m\beta_\gamma$
& $4.6\times 10^{-39} \,\beta_m\beta_\gamma$
& $6.3\times 10^{-44} \,\beta_\gamma^2$
&
& ---
&
\\
\end{tabular}
\end{ruledtabular}
\end{table*}

Multiplying Eq.~\eqref{eq:Dirac} by $-i(g^{00})^{-1}\underline\gamma^0$ on the left, we obtain the Schr\"odinger equation $i\partial_t\psi = H\psi$ with Hamiltonian
\begin{equation}
H = (- g^{00})^{-1}i\underline\gamma^0 m_e - (g^{00})^{-1}\underline\gamma^0\underline\gamma^i i D_i + i\omega_0 + e A_0.
\label{eq:HUnexpanded}
\end{equation}
To linear order in the perturbations, this can be written as
\begin{equation}
H = H_0 + H_\text{hfs} + \delta H_\text{fs} + \delta H_\text{hfs},
\end{equation}
where the unperturbed Hamiltonian for a one-electron atom is
\begin{subequations}
\begin{equation}
H_0 = m_e \beta + \alpha^i p_i + e A_0^{(0)},
\label{eq:H0}
\end{equation}
with momentum operator $p_i = -i\partial_i$. The matrices $\alpha^i = -\gamma^0\gamma^i$ and $\beta = i\gamma^0$ satisfy the algebra $\{ \alpha^i, \alpha^j \} = \delta^{ij}$, $\{ \alpha^i, \beta \} = 0$, and $\beta^2 = 1$. The coupling between the nuclear magnetic moment and the electron's angular momentum gives rise to a hyperfine structure and is treated as a small perturbation,
\begin{equation}
H_\text{hfs} = \gamma \epsilon_{ijk} \frac{\alpha^i I^j x^k}{r^3}.
\label{eq:hfs}
\end{equation}
The remaining perturbations $\delta H$ encapsulate gravitational effects. We categorize them as fine-structure and hyperfine-structure corrections, the former containing all the terms independent of the nuclear spin. These are
\begin{align}
\delta H_\text{fs} =&\, m_e \Phi_+ \beta + 2 \Phi_\text{N} \alpha^i p_i
\nonumber\\
&+ \frac{i}{2} \alpha^i (\partial_i \Phi_\text{N} - 3 \partial_i \Phi_\text{S}) + e \delta A_0,
\end{align}
whereas the hyperfine structure corrections are
\begin{equation}
\delta H_\text{hfs} = 2 \Phi_\text{N} \alpha^i e A_i^{(0)} + \alpha^i e \delta A_i.
\end{equation}
\end{subequations}
We note that the Hamiltonian receives other perturbations due to vacuum polarization, relativistic recoil, finite nuclear size effects, and so on~\cite{Codata,Tabulation}, which we will collectively refer to as ``QED corrections.'' As they contribute additively to the energy levels, there is no need to write them down explicitly, although they turn out to play an important role in Sec.~\ref{sec:EPviolations}.

So far, our derivation has kept both the chameleon and Newtonian potential for completeness. Making a rough estimate of their relative contributions, as we did in Sec.~\ref{sec:Background}, we find unsurprisingly that perturbations from the latter are unobservable. Specifically, we make the substitutions
\begin{equation}
x^i \sim (\zeta m_e)^{-1},\quad
p_i \sim \zeta m_e, \quad
\alpha^i \sim \zeta, \quad
\beta \sim 1,
\label{eq:EstimateSubstitutions}
\end{equation}
where $(\zeta m_e)^{-1}$ is of the order of the Bohr radius. With these replacements, the Newtonian terms all give a contribution of the schematic form $\sim G \zeta m_e^2 m_N$, which corresponds to a measly energy shift of $10^{-38}\,\text{eV}$ in hydrogen. It follows that for gravitational strength couplings $\beta_m$,~$\beta_\gamma \sim \mathcal O(1)$, the chameleon also causes energy shifts of a similar size. However, for strong couplings $\beta_m$,~$\beta_\gamma \gg 1$, this, in principle, may lead to observable effects. The remainder of this paper is concerned with such a possibility, and we will henceforth neglect the Newtonian potential terms.

\section{Discussion}
\label{sec:Discussion}

\subsection{Atomic precision test constraints}
\label{sec:atomictests}

Having included the effect of a chameleon-photon interaction sourcing corrections to the electromagnetic fields, our Hamiltonian differs from that in~\cite{BraxBurrage}. It is therefore worthwhile beginning our discussion by revisiting some of their analyses. The chameleon fine-structure corrections are
\begin{equation}
\delta H_\text{fs} \supset m_e \Phi_\text{S} \beta + \frac{3}{2} i \alpha^i \partial_i \Phi_\text{S} + e \delta A_0.
\label{eq:HChmFs}
\end{equation}
We have ignored the background gradient $\bar\phi_i$ for now, as this is constrained by torsion experiments~\cite{Cham1} to be negligible on Earth. Its effect in unscreened environments is considered in Sec.~\ref{sec:EPviolations}.

The second term in Eq.~\eqref{eq:HChmFs} is, roughly speaking, a coupling between the electron spin and the chameleon. For spherically symmetric profiles $\Phi_\text{S}$ as in Eq.~\eqref{eq:ChamSolution}, this term has a zero expectation value because of an exact cancellation between components of the electron wave function~\cite{Adkins} (further details are also given in the Appendix). However, this is no longer true if we relax any of the first three assumptions made in Sec.~\ref{sec:ChamProfile}, as these will generically break spherical symmetry. We do not consider such complications in this paper since we expect them to be subleading. Nevertheless, this term is interesting because it does not appear altogether if we work in the Einstein frame. This suggests an inequivalence between the two frames, and is discussed in Sec.~\ref{sec:Frames}.

For now, the remaining terms of interest give
\begin{align}
\langle \delta H_\text{fs} \rangle
\supset &
-2G \beta_m^2 m_e m_N \mathcal B_{-1} - \frac{1}{2} GZ\zeta \beta_m\beta_\gamma m_e \mathcal B_{-2}
\nonumber\\
& - G\zeta\beta_m\beta_\gamma m_N \mathcal A_{-2} - \frac{1}{6}GZ\zeta^2 \beta_\gamma^2 \mathcal A_{-3},
\label{eq:HChmFsExp}
\end{align}
where $\mathcal A_q = \langle r^q \rangle$ and $\mathcal B_q = \langle \beta r^q \rangle$ are radial expectation values. Explicit expressions are given in~\cite{Suslov} and are reproduced in the Appendix. Note that only the $\mathcal B$ terms were taken into account in~\cite{BraxBurrage}.

Let us define the additive correction to the energy of a transition line as $\delta E(\psi - \psi') = \langle\delta H \rangle(\psi) - \langle\delta H \rangle(\psi')$, for two states $\psi,\psi'$. Then the Ly\,$\alpha$ line has an extra correction $\delta E(2S_{1/2}-1S_{1/2}) \approx (10^{-38}\,\text{eV}) \beta_m^2 + \text{(smaller terms)}$ due to chameleon effects (see Table~\ref{table:fs} for the full expression). As was found in~\cite{BraxBurrage}, the dominant term, assuming $\beta_\gamma$ is not very much larger than $\beta_m$, comes from the mass of the nucleus sourcing a chameleon profile, which then couples to the electron mass via the metric. This is encapsulated in the $\mathcal B_{-1}$ term. Also worth mentioning is the $\mathcal A_{-3}$ term, which can be thought of as a chameleon-mediated backreaction of the electric field on itself. Containing a factor of $\beta_\gamma^2$, this term allows us to set an upper bound on $\beta_\gamma$, whereas previously only an upper bound on the product $\beta_m\beta_\gamma$ was possible.

\begin{figure}
\includegraphics[width=67.5mm]{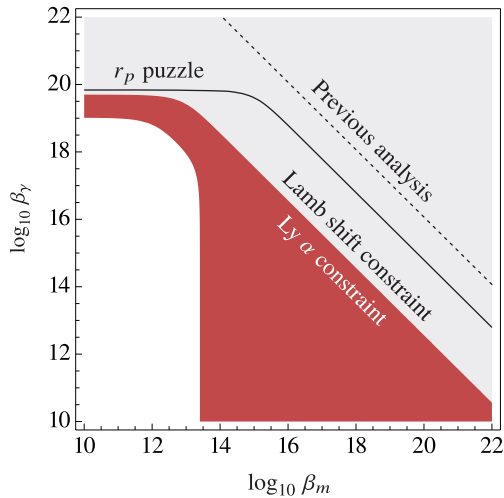}
\caption{The solid line gives the set of parameter values $(\beta_m,\beta_\gamma)$ for which a chameleonlike scalar is able to resolve the proton radius puzzle. The dotted line gives the corresponding set of values predicted by a previous analysis. Such a possibility is excluded by constraints (marked by shaded regions) from measurements of the (electronic) Lamb shift in hydrogen and the Ly\,$\alpha$ line. These constraints are much weaker than bounds placed by other methods (see text for details).}
\label{fig:constraints}
\end{figure}

We constrain these parameters as follows: Also provided in Table~\ref{table:fs} are the standard theoretical uncertainties for the energy level of each state $\sigma(\psi)_\text{th}$ as predicted by QED, and the standard uncertainty associated with the experimental measurement $\sigma(\psi-\psi')_\text{exp}$. For a conservative bound, we simply add these together to form a total standard uncertainty $\sigma(\psi-\psi')_\text{tot}$. For the Ly\,$\alpha$ line, we therefore require that the chameleon effects be smaller than $\sigma(2S_{1/2}-1S_{1/2})_\text{tot} = 1.1\times 10^{-11}\,\text{eV}$, which yields (see also Fig.~\ref{fig:constraints})
\begin{equation}
\log_{10}\beta_m \lesssim 13.4,\quad
\log_{10}\beta_\gamma \lesssim 19.0
\label{eq:ParameterBounds}
\end{equation}
at the 68\% confidence level. We have repeated this exercise for all hydrogen lines in Table~XVI of~\cite{Codata} with a relative standard uncertainty better than $10^{-10}$, concluding that the tightest bounds from hydrogen indeed come from the Ly\,$\alpha$ line. (We will later see in Sec.~\ref{sec:hfs} that the 21\,cm line has little constraining power.) It is worth commenting that our bound on $\beta_m$ is marginally better than what was found in~\cite{BraxBurrage} purely because we have used more recent values of $\sigma(\psi)$.

Limited by the theoretical uncertainty $\sigma(\psi)_\text{th}$, our updated constraints remain far from competitive and are unlikely to improve in the foreseeable future. The best bounds to date come from atom interferometry experiments~\cite{AI1,*AI2}, which are able to constrain $\log_{10}\beta_m \lesssim 5$~\footnote{All constraints, apart from those derived from atomic precision tests, are given at the 95\% confidence level.}. A different region of parameter space around $\beta_m \approx 1$ is also excluded from torsion experiments~\cite{EotWash}, although these apply only for a limited range of effective masses $m_\phi$ and can be evaded if the screening mechanism is adjusted (see~\cite{Compendium} for a review). Our constraint on $\beta_\gamma$ is similarly poor. The tightest bound again arises from torsion experiments~\cite{EotWash}, yielding $\log_{10}\beta_\gamma \lesssim 3$, but with the same caveats. A more universal bound from collider experiments~\cite{ColliderConstraints} gives $\log_\text{10}\beta_\gamma \lesssim 13.5$. The GammeV--CHASE experiment~\cite{GammeVCHASE} lowers this to $\log_\text{10}\beta_\gamma \lesssim 11$ for chameleons with effective mass $m_\phi \lesssim 1\,\text{meV}$.

\subsection{Regularizing singularities}
\label{sec:cutoff}

At this point, we must clarify a caveat to our calculations in Table~\ref{table:fs}. The $\mathcal A_{-3}$ integral formally diverges for $S_{1/2}$ and $P_{1/2}$ states, because their wave functions are nonzero at the origin, and a chameleon-induced correction to the Coulomb potential $ \propto r^{-3}$ is sufficiently steep. This is unphysical, and we naturally expect the finite size of the nucleus to regularize this divergence. For radii smaller than the nuclear charge radius $r_N$, the bare Coulomb potential $\propto r^{-1}$ is replaced by $A_0^{(0)} = - \zeta F(r)$, where $F$ is some form factor due to the nuclear charge distribution. As a crude example, $F = (2r_N)^{-1} (3 - r^2/r_N^2)$ for a uniformly charged sphere. In principle, this is calculable for any given model charge distribution, but we will remain agnostic about the exact form of $F$ in our argument. The delta function approximating the mass density of the nucleus must also be replaced by some distribution for $r \leq r_N$.

\begin{figure}
\includegraphics[width=66mm]{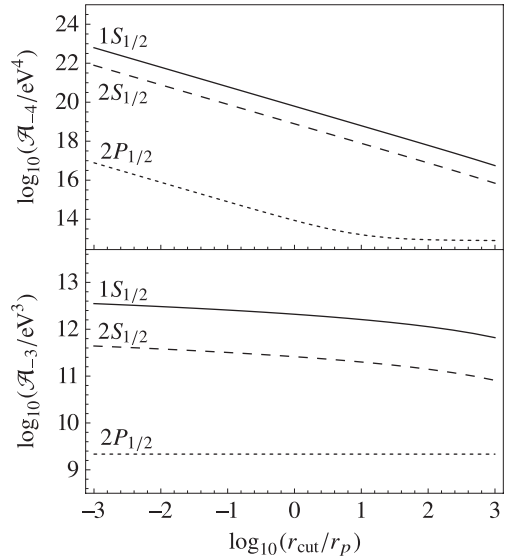}
\caption{Dependence of the regularized integrals ${\mathcal A}_{q}$ on the cutoff radius $r_\text{cut}$, normalized by the proton radius $r_p$, for the lowest-lying levels of hydrogen. The $q=-3$ integral is particularly insensitive to the cutoff.}
\label{fig:robust}
\end{figure}

As a result, the $r^q$ ($q<0$) dependence of each of the terms in the perturbation Hamiltonian becomes
\begin{equation}
r^q \to 
\begin{cases}
r^q & r > r_N,
\\
r_N^q \mathcal F(r) & r \leq r_N,
\end{cases}
\end{equation}
where $\mathcal F(r)$ is the corresponding form factor, which is generally different for each term in $\delta H$. Since the mass and charge distributions of the nucleus are now regular near the origin, so is $\mathcal F(r)$. Further assuming that $r^q$ and $r_N^q \mathcal F(r)$, and their first derivatives, match at the boundary $r = r_N$, this suggests $\mathcal F$ is bounded between $\mathcal F(r_N) =1 $ and $\mathcal F(0) > 1$, but of order unity. At leading order, we can approximate $\mathcal F(r) \approx \mathcal F(0)$.

Therefore, and only when necessary, we regularize the expectation values $\mathcal A_q$ using a cutoff radius,
\begin{equation}
\mathcal A_q(r_\text{cut}) =
\label{eq:Acutoff}
\int_0^{r_\text{cut}} \text{d}r \; |\psi|^2 r^2 r_\text{cut}^q 
+
\int_{r_\text{cut}}^\infty \text{d}r \; |\psi|^2 r^{2+q}.
\end{equation}
Rather than keep an explicit factor of $\mathcal F(0)$ in the first integral, we absorb its unknown value into $r_\text{cut}$, which we then allow to differ slightly from $r_N$. The $q=-3$ integral depends (almost) logarithmically on $r_\text{cut}$ (see Fig.~\ref{fig:robust}), suggesting that this term is not sensitive to the detailed nuclear structure and that our approximation captures the physics sufficiently well. We therefore set $r_\text{cut} = r_N \sim A^{1/3} r_p$ in this paper, where $A$ is the mass number and $r_p$ is the proton radius.

We further note that all the divergent terms in $\delta H$, once regularized, predict energy shifts that are many orders of magnitude smaller than those predicted by the other, finite terms. So even if one were uneasy about this procedure, it does not discount the predictive power of the finite terms. Given that we believe we have a good understanding of why and how these singularities should be removed, we have found it interesting to include their effects in our discussion.

\subsection{Lamb shift and the proton radius puzzle}
\label{sec:LambShift}

Both the $\mathcal A_{-2}$ and $\mathcal B_{-2}$ terms have a similar radial dependence, but they arise from different physics. The former is the mass of the nucleus sourcing a correction to the electric field, which explains the factor of $\zeta m_N$. The latter is the electric field sourcing a chameleon profile, which couples to the electron mass via the metric, thus picking up a factor of $Z\zeta m_e$. The inclusion of the $\mathcal A_{-2}$ term in this work explains why we find a much larger correction to the Lamb shift, on the order of $m_p/m_e$ for hydrogen.

It is interesting to ask if this larger contribution is able to resolve the proton radius puzzle. Stated simply, the puzzle is a $7\sigma$ discrepancy between measurements of the proton charge radius from experiments using electronic and muonic hydrogen. Electron-proton scattering and measurements of hydrogen and deuterium transition lines give $r_p = 0.8751(61)\,\text{fm}$~\cite{Codata}, whereas a measurement of the Lamb shift in muonic hydrogen~\cite{MuonPRP} gives $r_p = 0.84087(39)\,\text{fm}$. One approach to resolving this puzzle, and the one employed here, is to assume that the radius inferred from electronic hydrogen experiments is accurate. One then has to introduce ``new physics'' that gives a correction $\delta E(2P_{1/2}-2S_{1/2})(\mu^-) \approx 300\,\mu\text{eV}$~\cite{ReviewPRP} to resolve the discrepancy between the theoretical and experimental values of the muonic Lamb shift.

The solid line in Fig.~\ref{fig:constraints} shows the set of possible values of $\beta_m$ and $\beta_\gamma$ that allow the chameleon to resolve this puzzle, and is ruled out by constraints placed in Sec.~\ref{sec:atomictests}. The dotted line shows the same set of values inferred from the previous analysis in~\cite{BraxBurrage}. While the ratio of the $\mathcal A_{-2}$ to $\mathcal B_{-2}$ terms is of order $10^4$ in electronic hydrogen, this reduces to $m_p / m_\mu \sim 10$ in muonic hydrogen, explaining why the solid line is lowered only by an order of magnitude. Many other solutions have been proposed (see~\cite{ReviewPRP} for a review, and~\cite{Burgess} which introduces no new physics but proposes a change of boundary conditions at small $r$ on the electron wave function), although the puzzle still remains an open problem.

\subsection{Hyperfine splitting}
\label{sec:hfs}

We consider the effects of the nuclear spin in this section. We have neglected a magnetic contribution ${\bf B}_{(0)}^2$ as a source for the chameleon profile in Eq.~\eqref{eq:eomChamLinear}, as it is subleading with $\mathcal O(\gamma^2)$, whereas all terms here are linear in $\gamma$. Once again ignoring the background gradient $\bar\phi_i$, the Hamiltonian is
\begin{equation}
\delta H_\text{hfs} \supset \frac{\gamma G}{10}\epsilon_{ijk} \alpha^i I^j x^k
\left( 5\beta_m \beta_\gamma \frac{m_N}{r^4} + \beta_\gamma^2 \frac{Z\zeta}{r^5} \right).
\end{equation}
Both terms have operators of the form $\epsilon_{ijk} \alpha^i I^j x^k r^{-n}$, not unlike the hyperfine structure term in Eq.~\eqref{eq:hfs}, and hence can be treated using similar methods. Following the approach in~\cite{RoseElectron,*RoseAngMom}, we find
\begin{align}
\langle \delta H_\text{hfs} \rangle \supset 
-&\frac{\kappa\gamma G}{4\kappa^2 - 1} [ F(F+1) - I(I+1) - j(j+1) ]
\nonumber\\
&\times \left( \beta_m\beta_\gamma m_N \mathcal C_{-3} + \frac{Z\zeta}{5}\beta_\gamma^2 \mathcal C_{-4} \right),
\end{align}
where $\kappa$ is a quantum number, related to the angular-momentum quantum numbers $j$ and $l$ by Eq.~\eqref{eq:DefineKappa}; ${\bf F} = {\bf I} + {\bf j}$ is the total angular momentum of the system; and $\mathcal C_q = \langle i \alpha^r \beta r^q \rangle$ is a radial expectation value, with $\alpha^r$ denoting the projection of the Dirac $\alpha$ matrices along the radial direction. For $S_{1/2}$ and $P_{1/2}$ states, both $\mathcal C_{-3}$ and $\mathcal C_{-4}$ are singular, and we regularize just as we did for $\mathcal A_{-3}$ in Sec.~\ref{sec:cutoff}.

We find that the ground-state hyperfine transition picks up the correction
\begin{align}
\delta E(1S_{1/2}^{F=1}-1S_{1/2}^{F=0}) =&\,
(1.4\times 10^{-48} \,\text{eV}) \,\beta_m\beta_\gamma
\nonumber\\
&+ (6.5 \times 10^{-53 \pm 1}\,\text{eV})\beta_\gamma^2.
\end{align}
We have included a rough uncertainty in the second term because the $\mathcal C_{-4}$ integral depends on the cutoff as $\mathcal C_{-4} \sim r_\text{cut}^{-1}$ (this behavior is briefly explained in the Appendix). It follows that this term is more sensitive to the form factor $\mathcal F(r)$. We expect that the uncertainty due to our crude approximation can be at most $r_\text{cut} \sim 10^{0 \pm 1} r_p$, which is reflected in the above equation.

For comparison, the leading hyperfine-structure term [Eq.~\eqref{eq:hfs}] gives an energy splitting $\sim 10^{-6}\,\text{eV}$ in hydrogen. While measurements of this transition have been performed with great precision, achieving relative standard uncertainties at the level of $10^{-12}$~\cite{hfsMeasurement}, larger theoretical uncertainties make this transition  ill-suited for constraining parameters. This arises from a poor understanding of how the nucleus affects these states. However, even if theoretical predictions were able to match experimental precision, it is easy to check that this still produces weaker bounds on $\beta_m$ and $\beta_\gamma$ than in Eq.~\eqref{eq:ParameterBounds}.

\subsection{Equivalence principle violations}
\label{sec:EPviolations}

Our discussion so far has been concerning chameleon effects that might be present on Earth. In unscreened, astrophysical environments, the chameleon vev $\bar\phi$ and its gradient $\bar\phi_i$ can lead to violations of the Einstein equivalence principle (EP).

Recall that the effect of $\bar\phi$ is to induce effective gravitational and fine-structure constants [Eq.~\eqref{eq:VaryingConstants}]. If this has a value $\bar\phi_\oplus$ on Earth, we normalize $\Omega(\bar\phi_\oplus) = \varepsilon(\bar\phi_\oplus) = 1$, such that the bare constants $G_\textup{N}$ and $\alpha$ take their usual measured values. Deviations of $\Omega$ and $\varepsilon$ from unity give rise to a fractional change in all energy levels. For nonrelativistic states with $\zeta \ll 1$, the gross-structure energy levels are approximated by the modified Rydberg formula
\begin{equation}
E_n = -\frac{Z^2 \alpha^2 m_e}{2 \bar\varepsilon^2 n^2}.
\label{eq:RydbergEnergy}
\end{equation}
For $\bar\varepsilon$ sufficiently different from unity, this can be an observable effect. However, because this fractional change applies equally to all transition lines, distinguishing it from a cosmological redshift is difficult. In fact, it has long been understood that the spectra of many-electron atoms is far more sensitive to variations in the fine-structure constant. In particular, comparing absorption lines of Fe\,V and Ni\,V in the atmospheres of white dwarfs with those measured in the laboratory has been found to be a useful probe of this effect~\cite{WhiteDwarf,*WhiteDwarf2}.

We might wonder if the gradient $\bar\phi_i$ stands a better chance of detection in hydrogen. It leads to a fine-structure correction
\begin{equation}
\delta H_\text{fs} \supset \frac{1}{2} \zeta\beta_\gamma \bar\phi_i x^i r^{-1} = \frac{1}{2} \zeta\beta_\gamma |\nabla\bar\phi| \cos\theta,
\label{eq:Stark}
\end{equation}
where we have aligned the $z$~axis of the atom's rest frame with $\bar\phi_i$ in the second equality. This term is analogous to the Stark effect. As $\cos\theta$ is an odd-parity operator, the matrix element $\langle \psi'|\cos\theta|\psi \rangle$ vanishes unless the states $\psi,\psi'$ have opposite parity. This operator therefore mixes degenerate states with the same principal quantum number $n$ but with differing angular-momentum quantum numbers $\kappa = \pm|\kappa|$, such as states in the $S_{1/2}$ and $P_{1/2}$ levels. The energy splitting and the eigenstates can then be found using standard methods of degenerate perturbation theory. Unfortunately, QED corrections~\cite{Codata,Tabulation} much larger than this chameleon effect lift such degeneracies, nullifying any response to the presence of a chameleon gradient. The degeneracy between the $2S_{1/2}$ and $2P_{1/2}$ levels is famously broken by the Lamb shift, for instance.

How large must the background gradient be that its effect is not forbidden by QED corrections? If Eq.~\eqref{eq:Stark} is to be comparable to, or greater than, the Lamb shift $\sim 10^{-6}\,\text{eV}$, we require $\beta_\gamma |\nabla\bar\phi| \gtrsim Z^{-1}\,10^{-4}\,\text{eV} \approx Z^{-1}\,10^{19}\,g_\oplus,$ where we recall that $g_\oplus$ is the surface gravity on Earth. The best candidates for extreme surface gravities are neutron stars, but even they have surface gravities of ``only'' $\sim 10^{11}\,g_\oplus$. But this large surface gravity also implies that a chameleon will be screened by the thin-shell effect~\cite{Cham1}, so its gradient $|\nabla\bar\phi| \ll 10^{11} g_\oplus$.

The chameleon gradient also generates a hyperfine-structure correction
\begin{equation}
\delta H_\text{hfs} \supset
\frac{\gamma}{2}\beta_\gamma(\epsilon_{jkm}\delta_{il} - \epsilon_{ijk}\delta_{lm})\bar\phi^m \frac{\alpha^i I^j x^k x^l}{r^3}.
\label{eq:hfsGrad}
\end{equation}
The relevant matrix elements are between states with different total angular-momentum quantum numbers, but with identical electron quantum numbers $n$ and $\kappa$. For instance, the 21$\,$cm line from the ground-state splitting $1S_{1/2}^{F=1}-1S_{1/2}^{F=0}$ has $n=1,\kappa=-1$ in both states. Since the part of the operator in Eq.~\eqref{eq:hfsGrad} acting on the electron has odd parity, whereas the relevant matrix elements are between electron states of equal parity, this term also gives no contribution.

\subsection{Comparisons with the Einstein frame}
\label{sec:Frames}

When transforming to the Einstein frame, one conventionally makes the field redefinition $\tilde\psi = \Omega^{-3/2} \psi$~\cite{AnomCouplings}, where $\tilde\psi$ and $\psi$ are the spinors in the Einstein and Jordan frames, respectively. This recovers a canonical kinetic term for $\tilde\psi$, but because $\Omega$ multiplies the mass term in the Dirac action by one extra power, $\tilde\psi$ now has a chameleon-dependent mass $\tilde m_e = \Omega(\phi) m_e$. This result has occasionally been used to claim that the energy of Rydberg states [Eq.~\eqref{eq:RydbergEnergy}] also includes a factor of $\bar\Omega$. Indeed, had we repeated all of our calculations in the Einstein frame, we would find stray or missing factors of $\bar\Omega$ in our equations. This is a reflection of the fact that physical equivalence is preserved only if the units used to measure length, time, and energy also scale with appropriate factors of $\bar\Omega$ when transforming between the two frames~\cite{Faraoni}. Accounting for this ensures that no factor of $\bar\Omega$ is present in Eq.~\eqref{eq:RydbergEnergy}, as is to be expected, since energies are observable quantities that should be independent of the choice of frame.

There is a second subtlety at play here. We found in the Jordan frame that the chameleon couples to the electron mass via its effect on the metric. In the Einstein frame, the chameleon has no backreaction on the metric at linear order~\footnote{Assuming we continue to ignore the chameleon's effective mass.}, instead coupling directly to the mass. Consequently, the term $\frac{3}{2}i \alpha^i\partial_i \Phi_\text{S}$ is absent in the Einstein-frame Hamiltonian, suggesting that we have two different theories. The factor of $3/2$ makes it apparent that this discrepancy arises from the field redefinition. Strictly, field redefinitions do not alter the S-matrix, so they should not change the physical content of the theory. What has gone wrong is that our quantization procedure is inconsistent. In this paper, we have quantized $\psi$, treating everything else as a classical background, whereas the original analysis in~\cite{BraxBurrage} quantizes $\tilde\psi$ instead. We expect that physical equivalence is recovered if all fields are treated quantum mechanically. Practically, this is hard, leaving us with what is essentially a choice of interpretation: Either $\psi$ is interpreted as the electron and $\tilde\psi$ as a mixed electron-chameleon degree of freedom or vice versa. The field to quantize is the one we choose to call the electron. We would argue that $\psi$ is the more natural choice since it couples minimally to the Jordan frame metric, where it obeys properties typically ascribed to matter: Its trajectories follow geodesics, and it obeys the usual conservation laws $\nabla_\mu T^{\mu\nu} = 0$.

\section{Conclusions}
\label{sec:Conclusion}

The effects of a chameleon on the spectrum of one-electron atoms can be categorized into three broad classes: fine and hyperfine corrections due to physics in the atom's rest frame, the generation of EP-violating terms due to a background gradient in unscreened environments, and through the induction of effective constants of nature due to a local vev.

Previous studies~\cite{BraxBurrage} have already understood that the mass and electromagnetic fields of the nucleus generate a local chameleon profile, which then perturbs the Hamiltonian. We find that, additionally, the chameleon field acts as a secondary charge and current source (Sec.~\ref{sec:Maxwell}), leading to corrections to the bare electromagnetic fields. These give rise to perturbations in the Hamiltonian at the same order. Including this effect predicts a correction to the Lamb shift that is larger by a factor of $m_p/m_e$ in the case of hydrogen, prompting us to reconsider the prospects of resolving the proton radius puzzle with a chameleon (Sec.~\ref{sec:LambShift}). We quickly find that this is ruled out by constraints on the matter and photon couplings, $\beta_m$ and $\beta_\gamma$, obtained using precise measurements of hydrogen transition lines (Sec.~\ref{sec:atomictests}). The Ly\,$\alpha$ line remains the most stringent constraint, bounding $\log_{10}\beta_m \lesssim 13.4$ and $\log_{10}\beta_\gamma \lesssim 19.0$. These are universal to all screening mechanisms, provided only that the chameleon's effective mass in the laboratory satisfies $m_\phi \ll 4\,\text{keV}$. Although an improvement over~\cite{BraxBurrage}, these remain far from competitive and are unlikely to improve in the foreseeable future. (Tighter constraints obtained from other methods in the literature are discussed at the end of Sec.~\ref{sec:atomictests}, or see~\cite{Compendium} for a review.)

Including the effects of the nuclear spin (Sec.~\ref{sec:hfs}) allows for the study of hyperfine-structure corrections. As these terms contain steep potentials $\propto r^{-3}$ and $r^{-4}$, their expectation values formally diverge for $S_{1/2}$ and $P_{1/2}$ states, whose wave functions are nonzero at the origin. Such singularities are removed by finite nuclear size effects, which we have modeled by imposing a cutoff at the nuclear radius (see Sec.~\ref{sec:cutoff}). After regularization, we find that the predicted energy shifts are too small to place better bounds on $\beta_m$ and $\beta_\gamma$.

A background chameleon gradient generates EP-violating terms in the Hamiltonian, but simple parity arguments show that their expectation values must vanish (Sec.~\ref{sec:EPviolations}). Though disheartening in terms of observational prospects, it is interesting and somewhat surprising that, at linear order, one-electron atoms are blind to the presence of a background chameleon. While second-order terms will likely be parity-even and give a nonzero contribution, at this order we expect such terms to be too strongly suppressed to be useful.

Consequently, the most likely effect to be observed is also the simplest: The local vev of a chameleon, when coupled to the photon, gives rise to energy shifts by inducing an effective fine-structure constant $\alpha \to \alpha/\varepsilon(\phi)$. As this leads to the same fractional change in all energy levels, detecting variations in $\alpha$ between observer and source proves challenging since it would be indistinguishable from a cosmological redshift. Instead, we propose looking for this effect in settings where the fine-structure constant varies across the spatial extent of the source. As an example, it has recently been shown~\cite{RadialAcc} that the effect of cold dark matter on the radial acceleration relation of rotationally supported galaxies can be mimicked by a symmetron only partially screened on galactic scales. In such instances, the symmetron profile changes appreciably (at the percent level for a disk-dominated mass budget) between the center and outer regions of the galaxy. In fact, since the rotation curves for these galaxies~\cite{SPARC} are inferred from hydrogen spectroscopy, a symmetron-photon coupling would lead to an interesting interplay of effects. In the best-case scenario, this might further improve the agreement between theory and data. We hope to explore this in future work.

\begin{acknowledgments}
It is a pleasure to thank Clare Burrage for helpful discussions. This work was supported by Science and Technology Facilities Council United Kingdom Grants No. {ST/L000385/1} and No. {ST/L000636/1}. L.K.W. also acknowledges the support of the Cambridge Commonwealth, European and International Trust.
\end{acknowledgments}

\newpage
\appendix*
\section*{\uppercase{Appendix: Radial expectation values\\ of the hydrogen wave function}}
\setcounter{equation}{0}

The unperturbed Hamiltonian for a one-electron atom is given in Eq.~\eqref{eq:H0}, where the Dirac matrices take the form
\begin{equation}
\alpha^i=
\begin{pmatrix}
0 & \sigma^i \\
\sigma^i & 0
\end{pmatrix}, \quad
\beta =
\begin{pmatrix}
1 & 0 \\
0 & -1
\end{pmatrix}
\end{equation}
in the Dirac-Pauli representation, and $\sigma^i$ are the usual Pauli matrices. In this representation, we can write
\begin{equation}
\label{eq:Wavefunction}
\psi =
\begin{pmatrix}
g(r) \chi^m_\kappa \\
i f(r) \chi^m_{-\kappa}
\end{pmatrix}.
\end{equation}
The spinor spherical harmonics $\chi_\kappa^m$ are the eigenstates of the operators $-({\bf L}\cdot{\boldsymbol\sigma} + 1)$ and $J_z$, with corresponding quantum numbers $\kappa$ and $m$, respectively. Note that $\kappa$ is related to the usual angular-momentum quantum numbers $j$ and $l$ via
\begin{equation}
\label{eq:DefineKappa}
\kappa =
\begin{cases}
j + 1/2  & (l=j+1/2), \\
-(j+1/2) & (l=j-1/2).
\end{cases}
\end{equation}
Each electron state is therefore labeled by the three quantum numbers $\{ n,\kappa,m\}$, and has energy
\begin{equation}
E_{n\kappa m} = m_e \left( 1 + \frac{\zeta^2}{(n+\nu -|\kappa|)^2} \right)^{-1/2},
\end{equation}
where $\nu = \sqrt{\kappa^2 - \zeta^2}$. The radial functions $f$ and $g$ are real and depend only on the quantum numbers $n$ and $\kappa$.

In Sec.~\ref{sec:atomictests}, we have used the fact that $\langle \alpha^i \partial_i \Phi_\text{S} \rangle$ vanishes if $\Phi_\text{S}$ is spherically symmetric. This integral evaluates to~\cite{Adkins}
\begin{equation}
\langle{\alpha^i \partial_i h(r)}\rangle = \int_0^\infty\text{d}r\;(gf-fg)r^2 h'(r) = 0
\end{equation}
for any spherically symmetric function $h$, as claimed. The remaining expectation values in this paper have been defined in terms of three integrals~\cite{Adkins},
\begin{subequations}
\label{eq:DefineABC}
\begin{align}
\mathcal A_q &:= \langle r^q \rangle = \int_0^\infty \text{d}r\;(g^2 + f^2) r^{q+2},
\\
\mathcal B_q &:= \langle \beta r^q \rangle = \int_0^\infty \text{d}r\;(g^2 - f^2) r^{q+2},
\\
\mathcal C_q &:= \langle i \alpha^r \beta r^q \rangle = -2 \int_0^\infty \text{d}r\;gf \, r^{q+2}.
\end{align}
\end{subequations}
Defining $\mathcal E = E_{n\kappa m}/m_e$ and $ a = \sqrt{1-\mathcal E^2}$, direct evaluation yields~\cite{Suslov}
\begin{subequations}
\label{eq:AB}
\begin{align}
\mathcal B_{-1} &= \frac{a^2 m_e}{\zeta},
\\
\mathcal B_{-2} &= \frac{2 a^3 m_e^2 (2\nu^2 - \mathcal E \kappa)}{\zeta\nu(4\nu^2-1)},
\\
\mathcal A_{-2} &= \frac{2 a^3 m_e^2 \kappa(2 \mathcal E \kappa -1 )}{\zeta \nu(4\nu^2-1)},
\\
\mathcal A_{-3} &= \frac{2 a^3 m_e^3(3 \mathcal E^2 \kappa^2 -3 \mathcal E \kappa - \nu^2 + 1)}{\nu(\nu^2-1)(4\nu^2 -1)}.
\end{align}
\end{subequations}

As was pointed out in Sec.~\ref{sec:cutoff}, the $\mathcal A_{-3}$ integral diverges for $S_{1/2}$ and $P_{1/2}$ states, which have $|\kappa| = 1$. To see this, write $f$ and $g$ as series solutions~\cite{Adkins}
\begin{equation}
f(r) = r^{\nu-1} \sum_{k=0}^\infty f_k r^k, \quad
g(r) = r^{\nu-1} \sum_{k=0}^\infty g_k r^k,
\end{equation}
such that any of the three integrals has the form
\begin{equation}
\mathcal A_q = \int_0^\infty \text{d}r\; r^{2\nu + q} \sum_{k=0}^\infty a_k r^k,
\end{equation}
where $f_k$, $g_k$, and $a_k$ are appropriate coefficients. This diverges if $2\nu+q \leq -1$. For $|\kappa| = 1$, this happens when $q \leq -3$.

Obviously, Eqs.~\eqref{eq:AB} are no longer valid when these integrals diverge. Instead, we evaluate Eq.~\eqref{eq:DefineABC} directly while applying a cutoff radius $r_\text{cut}$, as defined in Eq.~\eqref{eq:Acutoff}. All the $q=-3$ integrals are particularly insensitive to the choice of cutoff, as is shown in Fig.~\ref{fig:robust} for $\mathcal A_q$. This is also easily seen. If the integrand is dominated by the singularity at the origin, then we can approximate
\begin{equation}
\mathcal A_q \sim \int_{r_\text{cut}}^\infty \text{d}r \; r^{2\nu + q} a_0 + \text{(subleading terms)}.
\end{equation}
For the choice $|\kappa|=1$ with $q = -3$, the exponent is $2\nu + q \approx -1.00005$. Occasionally, the integral is less sensitive to $r_\text{cut}$ than this would predict (e.g., the $2P_{1/2}$ lines in Fig.~\ref{fig:robust}). This occurs when there is also a finite but substantial contribution from the integrand away from $r=0$.

\bibliography{h.bib}
\end{document}